\documentclass[12pt,preprint]{aastex}

\def\lax{{$\mathrel{\hbox{\rlap{\hbox{\lower4pt\hbox{$\sim$}}}\hbox{$<$}}}$}}
\def\gax{{$\mathrel{\hbox{\rlap{\hbox{\lower4pt\hbox{$\sim$}}}\hbox{$>$}}}$}}

\usepackage{color}

\def\vec#1{\mbox{\boldmath $#1$}}
\usepackage{lscape}

\slugcomment{}

\begin{document}
\title{Warped Circumbinary Disks in Active Galactic Nuclei}

 \author{
 Kimitake \textsc{Hayasaki}\altaffilmark{1},
 Bong Won \textsc{Sohn}\altaffilmark{1,2},
Atsuo T. \textsc{Okazaki}\altaffilmark{3},
Taehyun \textsc{Jung}\altaffilmark{1},
Guangyao \textsc{Zhao}\altaffilmark{1},
   and
 Tsuguya \textsc{Naito}\altaffilmark{4}
   }
\altaffiltext{1}{Korea Astronomy and Space Science Institute, Daedeokdaero 776, Yuseong, Daejeon 305-348, Korea}
\email{kimi@kasi.re.kr}
\altaffiltext{2}{Department of Astronomy and Space Science, University of Science and Technology, 217 Gajeong-ro,
Daejeon, Korea}
\altaffiltext{3}{Faculty of Engineering, Hokkai-Gakuen University, Toyohira-ku, Sapporo 062-8605, Japan}
\altaffiltext{4}{Faculty of Management Information, Yamanashi Gakuin University, Kofu, Yamanashi 400-8575, Japan}
%
\begin{abstract}
We study a warping instability of a geometrically thin, non-self-gravitating disk surrounding 
binary supermassive black holes on a circular orbit. Such a circumbinary disk is subject to 
not only tidal torques due to the binary gravitational potential but also radiative torques due 
to radiation emitted from an accretion disk around each black hole. We find that {a} circumbinary 
disk initially aligned with the binary orbital plane is unstable to radiation-driven warping beyond 
the marginally stable warping radius, which is sensitive to both the ratio of vertical to horizontal 
shear viscosities and the mass-to-energy conversion efficiency. As expected, the tidal torques 
give no contribution to the growth of warping modes but tend to align the circumbinary disk with 
the orbital plane. Since the tidal torques can suppress the warping modes in the inner part of 
circumbinary disk, the circumbinary disk starts to be warped {at radii larger than} the 
marginally stable warping radius. If the warping radius is of the order of $0.1\,\rm{pc}$, a resultant 
semi-major axis is estimated to be of the order of $10^{-2}\,\rm{pc}$ to $10^{-4}\,\rm{pc}$ for 
$10^{7}{\rm{M}}_\odot$ black hole. We also discuss {the} possibility that the central object{s} of observed warped maser disks in {active galactic nuclei are} binary supermassive black holes with a triple disk: two accretion disks around {the individual} black holes and one circumbinary disk surrounding them.
%
\end{abstract}
%
\section{Introduction}
%

%
%
There is {strong} evidence that most galaxies harbor supermassive black holes (SMBHs) with mass $10^5{\rm{M}}_\odot\lesssim{M}\lesssim10^{10}{\rm{M}}_\odot$ at their centers \citep{kr95}. Hitherto, SMBHs have been found in 87 galaxies by observing {the} proper motion of stars bound by the SMBHs or by detecting radiation emitted from gas pulled gravitationally by the SMBHs \citep{kho13}. $\rm{H}_2\rm{O}$ maser emission from active galactic nuclei (AGNs) in spiral galaxies provides {a strong tool} to measure SMBH {masses}, because it shows a rotating disk on a subparsec scale with a nearly Keplerian velocity distribution around the SMBH. Those disks, so-called maser disks, have been observed at the centers of NGC\,4258 \citep{miyoshi95}, NGC\,1068 \citep{green97}, NGC\,3079 \citep{yama04}, the Circinus galaxy \citep{green03}, UGC\,3789 \citep{reid09}, NGC\,6323 \citep{braatz07}, NGC\,2273, NGC\,6264, and some more objects \citep{kuo11}.

%
%
Several maser disks show {warped structure} at the radii of the order of 
$0.1\,\rm{pc}$ \citep{green03,he05,kond08,kuo11}. 
{From an} observational point of view, maser spots on the disk in NGC~4258 
are spatially distributed along a line on each side of a central black hole. 
The SMBH is then thought to be located at the center of a line connecting those two lines 
by a simple extrapolation, and the disk starts to {warp} at the innermost maser spot.
What mechanism makes the disk warped still remains an open question.

%
Several promising theoretical ideas have been proposed for explaining disk warping.
\cite{pr96} showed that centrally illuminated accretion disks are unstable to warping 
{due to} the reaction force of reradiated radiation. Such a radiation-driven warping 
mechanism has also been applied to explain the disk warping in the context of X-ray 
binaries \citep{mb97,wijers99,martin07,martin09}. If angular momentum vector of an 
accretion disk around a spinning black hole is misaligned with the spin axis, 
{differential} Lense-Thirring torque due to the frame-dragging effect {aligns} 
the inner part of the disk with the black-hole equatorial plane. 
Since the outer part of the disk {retains its initial orientation,} 
the resultant disk is warped \citep{bp75}.
This Bardeen-Petterson effect is also considered to be a plausible mechanism for disk 
war{p}ing in maser disks \citep{caproni07}. Moreover, \cite{balex09} proposed that the warped 
disk at the center of NGC 4258 is caused by the process of resonant relaxation, which is 
a rapid relaxation mechanism to exchange angular momentum between the disk and the 
stars moving under the nearly spherical potential dominated by the SMBH. These mechanisms 
have been discussed based on the assumption that the central object surrounded by the warped 
maser disk is a single SMBH.

%
%
The tight correlation between the mass of SMBHs and the mass or luminosity of the bulge of their 
host galaxies strongly support the idea that SMBHs have grown with the growth of their host galaxies 
(\citealt{mtr98,geb00,fm00,mm13}; see also \citealt{kho13} for a review). This relationship suggests that 
each SMBH at the center of each galaxy should have evolved toward coalescence in a merged galaxy. 
If this is the case, a binary of SMBHs on a subparsec scale or less should be formed in a merged galactic 
nucleus before two black holes finally coalesce, yet no binary SMBHs have clearly been identified so far {despite} some claims (see \citealt{ko06,po12} for reviews and references therein).

%
%
In the standard scenario of evolution of merging black holes \citep{bege80}, 
{It is still unclear} what mechanism efficiently {extracts} the orbital angular momentum 
of binary SMBHs on subparsec scales within a Hubble time. 
Once the binary orbit decays down to {sub-milliparsec radii}, the binary rapidly 
coalesces by gravitational wave emission (e.g., \citealt{sjd13}). Therefore, binary SMBHs on sub-parsec 
to sub-milliparsec scales are a "missing link'' in the merger history of SMBHs. Since the size of warped 
maser disks is identified to be of the order of $0.1\,\rm{pc}$, the binary separation should be smaller than 
this scale if the central object of warped maser disks is binary SMBHs. In oder to understand evolution 
of binary SMBHs, it is thus important to study a possible link between the presence of binary SMBHs and 
the warping of maser disks.

%
%
In this paper, we discuss the possibility that the central object{s of warped maser disks are} a potential candidate for binary SMBHs {with} a $\lesssim0.1\rm{pc}$. In section~\ref{sec:2}, we {describe the} external torques acting on a circumbinary disk. {We consider both} the tidal torques {originating} from a binary potential and the radiative torques due to radiation emitted from two inner circum-black-hole disks (accretion disks). In section~\ref{sec:3}, we study the evolution of a slightly tilted circumbinary disk subject to those two torques. In section~\ref{sec:4}, we apply our model to observed warped maser disks in AGNs and then estimate the semi-major axis of binary SMBHs. Finally, section~\ref{sec:5} is devoted to summary and discussion of our scenario.

%
\section{External Torques acting on the circumbinary disk}
\label{sec:2}
%

\begin{figure}[ht!]
\begin{center}
\includegraphics[width=10cm]{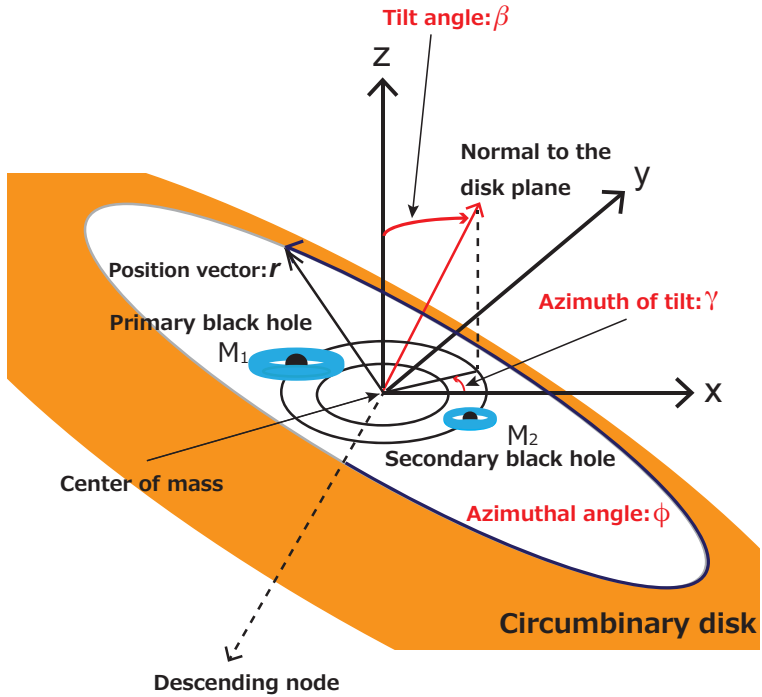}
\end{center}
\caption{
Configuration of a triple disk system composed of two accretion disks around {the individual} black holes and a circumbinary disk surrounding them. There are two angles ($\beta, \gamma$) which specify the orientation of the circumbinary disk plane with respect to the binary orbital plane ($x$-$y$ plane). The azimuthal angle ($\phi$) of an arbitrary position on the circumbinary disk is measured from the descending node.
}
\label{fig:schmatic}
\end{figure}

Let us consider the torques from the binary potential 
acting on the circumbinary disk {surrounding} the binary on a circular orbit.
Figure~\ref{fig:schmatic} illustrates a schematic picture of our model; 
binary black holes orbiting each other are surrounded by
a misaligned circumbinary disk.
The binary is put on the $x$-$y$ plane with its center of mass being at the origin 
in the Cartesian coordinate.
The masses of the primary and secondary black holes are represented 
by $M_1$ and $M_2$, respectively, and $M=M_1+M_2$. We put a circumbinary disk around the origin.
The unit vector of specific angular momentum of the circumbinary disk is expressed 
by (e.g. \citealt{pr96})
\begin{equation}
\mbox{\boldmath $l$} = \cos\gamma\sin\beta\vec{i} + \sin\gamma\sin\beta\vec{j} +\cos\beta\vec{k},
\label{eq:damvec}
\end{equation}
where $\beta$ is the tilt angle between the circumbinary disk plane and the 
binary orbital plane, and $\gamma$ is the azimuth of tilt. 
Here, $\vec{i}$, $\vec{j}$, and $\vec{k}$ are unit vectors in the $x$, $y$, and $z$, respectively.
The position vector of the circumbinary disk can be expressed by
\begin{equation}
\mbox{\boldmath $r$}=r(\cos\phi\sin\gamma+\sin\phi\cos\gamma\cos\beta)\vec{i}
+ r(\sin\phi\sin\gamma\cos\beta-\cos\phi\cos\gamma)\vec{j} 
-r\sin\phi\sin\beta\vec{k}
\label{eq:rin}
\end{equation}
where the azimuthal angle $\phi$ is measured from the descending node. 
The position vector of each black hole is given by
\begin{equation}
\mbox{\boldmath $r$}_{i}=r_{i}\cos{f_i}\vec{i}+r_{i}\sin{f_i}\vec{j} \hspace{2mm}(i=1,2),
\label{eq:ri}
\end{equation}
where $r_{i}=\xi_{i}a$ with $\xi_1\equiv q/(1+q)$ and $\xi_2\equiv 1/(1+q)$. 
Here, $q=M_2/M_1$ is the binary mass ratio and $a$ is the semi-major axis of the binary.
These and other model parameters are listed in Table~1.

%
%
\begin{table}[!ht]
  \caption{Model parameters}
     \begin{tabular}{ll}
       \hline
       Definition & Symbol\\
       \hline
       Total black hole mass  & $M$  \\
       Primary black hole mass & $M_1$ \\
       Secondary black hole mass & $M_2$ \\
       Schwarzschild radius & $r_{\rm{S}}=2GM/c^2$ \\
       Binary mass ratio & $q=M_2/M_1$ \\
       Mass ratio parameters & $\xi_1=q/(1+q)$, $\xi_2=1/(1+q)$ \\
       Binary semi-major axis & $a$ \\ 
       Orbital frequency & $\Omega_{\rm{orb}}=\sqrt{GM/a^3}$ \\
       Orbital period & $P_{\rm{orb}}=2\pi/\Omega_{\rm{orb}}$ \\
       Tilt angle & $\beta$ \\
       Azimuth of tilt & $\gamma$ \\
       Azimuthal angle & $\phi$ \\
       Shakura-Sunyaev viscosity parameter & $\alpha$ \\
       Horizontal shear viscosity & $\nu_1$ \\
       Vertical shear viscosity & $\nu_2$\\
       Ratio of vertical to horizontal shear viscosities & $\eta=\nu_2/\nu_1$\\
       Mass-to-energy conversion efficiency & $\epsilon$ \\
       Luminosities emitted from two accretion disks & $L_1$, $L_2$ \\
       Total luminosity & $L=L_1+L_2$ \\ 
       Binary irradiation parameter & $\zeta=(\xi_1^2L_1+\xi_2^2L_2)/L$ \\
       \hline
     \end{tabular}
   \label{tb:t1}
\end{table}

%
%

%
\subsection{Gravitational Torques}
%

The gravitational force on the unit mass at position $\vec{r}$ on the circumbinary disk can be written by
\begin{eqnarray}
\vec{F}_{\rm{grav}}=-\sum_{i=1}^{2}\frac{GM_i}{|\vec{r}-\vec{r}_i|^3}(\vec{r}-\vec{r}_i)
\end{eqnarray}
The corresponding torque is given by
\begin{eqnarray}
\vec{t}_{\rm{grav}}=\vec{r}\times\vec{F}_{\rm{grav}}=\sum_{i=1}^2\frac{GM_i}{|\vec{r}-\vec{r}_i|^3}(\vec{r}\times\vec{r}_i)
\end{eqnarray}
We consider the tidal warping/precession with timescales much longer than 
local rotation period of the circumbinary disk. This allows us to use the torque 
averaged in the azimuthal direction and over the orbital period:
\begin{eqnarray}
\langle\vec{T_{\rm{grav}}}\rangle
=
\frac{1}{4\pi^2}\int_0^{2\pi}
\int_0^{2\pi}
\frac{
\Sigma
}
{|{\vec{J}}|}\vec{t}_{\rm{grav}}
\,d\phi
d(\Omega_{\rm{orb}}t)
=\frac{3}{8}\xi_1\xi_2\Omega\left(\frac{a}{r}\right)^2
\Biggr[
\sin\gamma\sin2\beta\vec{i}
-
\cos\gamma\sin2\beta\vec{j}
\Biggr],
\label{eq:tgrav}
\end{eqnarray}
where ${\vec{J}}\equiv{r}^2\Omega\Sigma\vec{l}$, and $\Omega_{\rm{orb}}=\sqrt{GM/a^3}$ 
and $\Omega=\sqrt{GM/r^3}$ are the angular frequencies of binary motion and mean motion of circumbinary disk at $r$, respectively. Here, we used for the integration the following approximations:
\begin{eqnarray}
|\vec{r}-\vec{r}_i|^{-3}&\approx&r^{-3}\left[1+3\frac{\vec{r}\cdot\vec{r}_i}{r^2}+\mathcal{O}((r_i/r)^2)\right]
\nonumber 
\end{eqnarray}
For a small tilt angle $\beta\ll1$, equation~(\ref{eq:tgrav}) is reduced to
\begin{eqnarray}
\langle\vec{T}_{\rm{grav}}\rangle
\approx
\frac{3}{4}
\xi_1\xi_2
\Omega
\left(\frac{a}{r}\right)^2
\Biggr[
l_y
\vec{i}
-
l_x
\vec{j}
\Biggr],
\label{eq:tgrav2}
\end{eqnarray}
where $l_x$ and $l_y$ can be written from equation~(\ref{eq:damvec}) as $l_x=\beta\cos\gamma$ and $l_y=\beta\sin\gamma$.
 
The tidal torques tend to align the tilted circumbinary disk with the orbital plane (c.f. \citealt{bate00}). 
For $e=0$, such a tidal alignment timescale is given by
{
\begin{eqnarray}
\tau_{\rm{tid}}=\frac{\sin\beta}{|\langle\vec{T}_{\rm{grav}}\rangle|}\approx\frac{8}{3\pi}
\left(\frac{1/4}{\xi_1\xi_2}\right)
\left(\frac{r}{a}\right)^{7/2}
P_{\rm{orb}},
\label{eq:ttid}
\end{eqnarray}
}
where $P_{\rm{orb}}=2\pi/\Omega_{\rm{orb}}$ is the binary orbital period.
Since the inner edge of the circumbinary disk is estimated to be 
$\sim1.7a$ \citep{al94}, the tidal alignment timescale is longer than 
the binary orbital period.
 
%
%
\subsection{Radiative Torques}
%
%

%
%
If there is an accretion disk around each black hole, the circumbinary 
disk can be illuminated by light emitted from each accretion disk. 
The re-radiation from the circumbinary-disk surface, which absorbs 
photons emitted from these accretion disks, causes {a} reaction force. 
This is the origin of the radiative torques. {Below we take two accretion 
disks as point irradiation sources, because their sizes are much smaller 
than that of the circumbinary disk. Note that negligible contribution arises 
from other radiation sources such as an accretion stream from the circumbinary 
disk towards each accretion disk \citep{hms07,mm08,rdscc11,dza13} and an 
inner rim of the circumbinary disk, because the mass-to-energy conversion 
efficiencies in these regions are negligible in comparison with those in the 
inner parts of the accretion disks. Furthermore, $\sin\beta$ would be larger 
than the dimensionless scale-height of each accretion disk, which is typically of 
the order of 0.01. If not, the radiation from the inner parts of the accretion disks 
is shadowed and less flux will reach the circumbinary disk, except for the case 
that the circumbinary disk is flaring.
}


%
%
Since the surface element on the circumbinary disk is given in the polar 
coordinates by
\begin{eqnarray}
\vec{dS}
=\frac{\partial\vec{r}}{\partial{r}}\times\frac{\partial\vec{r}}{\partial\phi}\,\,drd\phi
=
\left[
\vec{l}-\vec{r}
\left(
-\frac{\partial\beta}{\partial{r}}\sin\phi
+\frac{\partial\gamma}{\partial{r}}\cos\phi\sin\beta
\right)
\right]
\,rdrd\phi,
\end{eqnarray}
the radiative flux at $d\vec{S}$ is given by
\begin{eqnarray}
dL=dL_1+dL_2=\frac{1}{4\pi}\sum_{i=1}^{2}\frac{L_i}{|\vec{r}-\vec{r}_i|^2}
\frac{|(\vec{r}-\vec{r}_i)\cdot{d}\vec{S}|}{|\vec{r}-\vec{r}_i|},
\end{eqnarray}
where $L$ is the sum of the luminosity of the radiation emitted from the primary black hole, $L_1$,
and that from the secondary black hole, $L_2$. Here, we assume that the surface element is not shadowed by other interior parts of the circumbinary disk. If we ignore limb darkening, the force acting on the disk surface by the radiation reaction has the magnitude of $(2/3)(dL/c)$ and is antiparallel to the local disk normal \citep{pr96}.
The total radiative force on $d\vec{S}$ can then be written by
\begin{eqnarray}
d\vec{F}_{\rm{rad}}
=
\frac{1}{6\pi{c}}
\sum_{i=1}^{2}L_i
\frac{|(\vec{r}-\vec{r}_i)\cdot
d\vec{S}|}{|\vec{r}-\vec{r}_i|^3}
\frac{d\vec{S}}{|d\vec{S}|}.
\end{eqnarray}
Consequently, the total radiative torque acting on a ring of radial width $dr$ is given by
\begin{eqnarray}
d\vec{T}_{\rm{rad}}
&=&
\oint\vec{r}\times{d\vec{F}_{\rm{rad}}}=
\frac{1}{6\pi{c}}
\oint
\sum_{i=1}^{2}
\Biggr[
L_i
\frac{|(\vec{r}-\vec{r}_i)\cdot
d\vec{S}|}{|\vec{r}-\vec{r}_i|^3}
\Biggr]
\frac{\vec{r}\times{d}\vec{S}}{|d\vec{S}|}
\nonumber \\
&\approx&
\frac{L}{6\pi{c}}
\frac{1}{r^3}
\oint
|
\vec{r}\cdot{d}\vec{S}
|
\frac{\vec{r}\times{d}\vec{S}}{|d\vec{S}|}
+
\frac{1}{2\pi{c}}
\frac{1}{r^5}
\oint
\sum_{i=1}^{2}
L_i
|
(\vec{r}-\vec{r}_i)\cdot{d}\vec{S}
|
(\vec{r}\cdot\vec{r}_{i})
\frac{\vec{r}\times{d}\vec{S}}{|d\vec{S}|},
\label{eq:radtorque}
\end{eqnarray}
where  $\oint|(\vec{r}-\vec{r}_i)\cdot{d}\vec{S}|(\vec{r}\times{d}\vec{S})/|d\vec{S}|=\oint|\vec{r}\cdot{d}\vec{S}|(\vec{r}\times{d}\vec{S})/|d\vec{S}|$ holds for $\beta\ll1$ and ${r}\partial{\beta}/\partial{r}\ll1$, 
and the first term, which we call $d\vec{T}_0$, of the right-hand side of equation~(\ref{eq:radtorque}) corresponds to equation (2.15) of \cite{pr96}:
\begin{eqnarray}
d\vec{T}_0=\frac{L}{6c}\left(r\frac{\partial{l_y}}{\partial{r}}\vec{i}-r\frac{\partial{l_x}}{\partial{r}}\vec{j}
\right)dr
\label{eq:dt0}
\end{eqnarray}
and the second term, which we call $d\vec{T}_{\rm{orb}}$, is originated from the orbital motion of the binary.

%
%
Here, we consider the radiation-driven warping/precession with timescales much longer than 
the orbital period, as in the case of tidally driven warping/precession.
The orbit-average of the torque $d\vec{T}_{\rm{rad}}$ is then given by
\begin{eqnarray}
\langle
d\vec{T}_{\rm{rad}}
\rangle
&=&
\frac{1}{2\pi}\int_{0}^{2\pi}
d\vec{T}_{\rm{rad}}\,d(\Omega_{\rm{orb}}t)
\approx
{d}\vec{T}_0+\frac{1}{2\pi}\int_{0}^{2\pi}d\vec{T}_{\rm{orb}}\,d(\Omega_{\rm{orb}}t)
\nonumber \\
&=&
\frac{L}{6c}
\Biggr\{
\Biggr(
-\frac{3}{2}\zeta\left(\frac{a}{r}\right)^2
{l_y}
+r\left[1-\frac{3}{2}\zeta\left(\frac{a}{r}\right)^2
\right]\frac{\partial{l_y}}{\partial{r}}
\Biggr)\vec{i}
\nonumber \\
&+&
\Biggr(
\frac{3}{2}\zeta\left(\frac{a}{r}\right)^2
{l_x}
-r\left[1-\frac{3}{2}\zeta\left(\frac{a}{r}\right)^2
\right]\frac{\partial{l_x}}{\partial{r}}
\Biggr)\vec{j}
\Biggr\}dr,
\label{eq:radt}
\end{eqnarray}
where $\zeta\equiv\xi_1^2L_1+\xi_2^2L_2/L$ is a binary irradiation parameter.
Note that $\zeta\lesssim1$: $\zeta\rightarrow{L_2/L}\lesssim1$ in the case of  $q\rightarrow0$ and $\zeta=1/4$ 
in the case of $q=1$. For $r\gg{a}$ or $\zeta=0$, $\langle{d\vec{T}_{\rm{rad}}}\rangle$ is reduced to $d\vec{T}_0$.

%
%
From equation~(\ref{eq:radt}), the specific radiative torque averaged over azimuthal angle and orbital phase 
is given by
\begin{eqnarray}
 \langle\vec{T}_{\rm{rad}}\rangle
 &=&
\frac{1}{|{\vec{J}}|}\frac{1}{2\pi{r}}\frac{\langle{d}\vec{T}_{\rm{rad}}\rangle}{dr}
=
\frac{\Gamma}{r}
\Biggr\{
\Biggr(
-\frac{3}{2}\zeta\left(\frac{a}{r}\right)^2
{l_y}
+r\left[1-\frac{3}{2}\zeta\left(\frac{a}{r}\right)^2
\right]\frac{\partial{l_y}}{\partial{r}}
\Biggr)\vec{i}
\nonumber \\
&+&
\Biggr(
\frac{3}{2}\zeta\left(\frac{a}{r}\right)^2
{l_x}
-r\left[1-\frac{3}{2}\zeta\left(\frac{a}{r}\right)^2
\right]\frac{\partial{l_x}}{\partial{r}}
\Biggr)\vec{j}
\Biggr\},
\end{eqnarray}
where $\Gamma=L/(12\pi\Sigma{r^2}\Omega{c})$. Assuming that $L=\epsilon\dot{M}c^2$ with 
the mass-to-energy conversion efficiency $\epsilon$ and {the mass accretion rate of the circumbinary disk  $\dot{M}=3\pi\nu_1\Sigma$}, then the growth timescale of {a} warping mode induced by the radiative torque, $r/\Gamma$, can be estimated to be
\begin{eqnarray}
\tau_{\rm{rad}}
=\frac{4}{\epsilon\alpha}\left(\frac{H}{r}\right)^{-2}\frac{r}{c}
\sim5\times10^{5}
\left(\frac{0.1}{\alpha}\right)\left(\frac{0.1}{\epsilon}\right)\left(\frac{H/r}{0.01}\right)^{-2}
\left(\frac{r}{r_{\rm{S}}}\right)^{-1/2}
\left(\frac{r}{a}\right)^{3/2}
P_{\rm{orb}},
\label{eq:trad}
\end{eqnarray}
where $\nu_1=\alpha{c}_{\rm s}{H}$ is the shear viscosity of the disk with the Shakura-Sunyaev viscosity parameter $\alpha$, $c_{\rm{s}}$ is the sound speed, and $H$ is the scale-height of the circumbinary disk. Here, $\epsilon\approx0.1$ is adopted for a Schwarzschild black hole case and $\epsilon=0.42$ for an extreme Kerr black hole case (e.g., see \citealt{kato08}). Since it is clear that $\tau_{\rm{rad}}\gg{P}_{\rm{orb}}$ for a geometrically thin disk, our assumption for the orbit-averaged radiative torque is ensured.

%
\section{Tilt angle evolution of circumbinary disks}
\label{sec:3}
%

%
%
In this section, we investigate the response of the circumbinary disk, which is initially aligned 
with the orbital plane, for external forces.
The mass conservation equation is given by
\begin{equation}
\frac{\partial\Sigma}{\partial{t}}+\frac{1}{r}\frac{\partial}{\partial{r}}(r\Sigma{v}_{r})=0,
\label{eq:mass}
\end{equation}
where $v_{r}$ is the radial velocity.
The angular momentum equation is given by\citep{pp83}
\begin{equation}
\frac{\partial{\vec{J}}}{\partial{t}}+\frac{1}{r}\frac{\partial}{\partial{r}}(r{v}_{r}{\vec{J}})=\frac{1}{r}\frac{\partial\vec{G}_{\rm{vis}}}{\partial{r}}
+|{\vec{J}}|\vec{T}_{\rm{ex}},
\label{eq:angmom}
\end{equation}
where $\vec{G}_{\rm{vis}}$ {represents} the viscous torques of the circumbinary disk.

%
%
The external torque $\vec{T}_{\rm{ex}}$ is written as the sum of tidal torques and radiative torques,
\begin{eqnarray}
\vec{T}_{\rm{ex}}
&=&\langle\vec{T}_{\rm{grav}}\rangle + \langle\vec{T}_{\rm{rad}}\rangle
\nonumber \\
&=&
\Biggr\{
\frac{3}{2}\left(\frac{a}{r}\right)^2
\left[
\frac{1}{2}\xi_1\xi_2
\Omega
-\frac{\zeta}{\tau_{\rm{rad}}}
\right]l_y
+\Gamma\left[1-\frac{3}{2}\zeta\left(\frac{a}{r}\right)^2
\right]\frac{\partial{l_y}}{\partial{r}}
\Biggr\}\vec{i}
\nonumber \\
&-&
\Biggr\{
\frac{3}{2}\left(\frac{a}{r}\right)^2
\left[
\frac{1}{2}
\xi_1\xi2
\Omega
-\frac{\zeta}{\tau_{\rm{rad}}}
\right]l_x
+\Gamma\left[1-\frac{3}{2}\zeta\left(\frac{a}{r}\right)^2
\right]\frac{\partial{l_x}}{\partial{r}}
\Biggr\}\vec{j}.
\nonumber \\
\end{eqnarray}
The evolution equation for disk tilt is obtained from equation (\ref{eq:mass}) and (\ref{eq:angmom}) as
\begin{equation}
\frac{\partial\vec{l}}{\partial{t}}
+
\left[
v_r-\nu_1\frac{\Omega^{'}}{\Omega}
-\frac{1}{2}\nu_2\frac{(r^3\Omega\Sigma)^{'}}{r^3\Omega\Sigma}
\right]
\frac{
\partial{\vec{l}}
}{
\partial{r}
}
=\frac{\partial}{\partial{r}}\left(\frac{1}{2}\nu_2\frac{\partial{\vec{l}}}{\partial{r}}\right)
+\frac{1}{2}\nu_2\left|\frac{\partial\vec{l}}{\partial{r}}\right|^2\vec{l}+\vec{T}_{\rm{ex}}
\label{eq:dtilt}
\end{equation}
\citep{pr96}, where $\nu_1$ and $\nu_2$ are respectively the horizontal shear viscosity and the vertical 
shear viscosity, the latter of which is associated with reducing disk tilt. The primes indicate differentiation 
with respect to $r$. For simplicity, we adopted the same assumptions for the circumbinary disk structure 
as in \cite{pr96} that $v_r=\nu_1\Omega^{'}/\Omega$, $r^3\Omega\Sigma$ is constant, and $\nu_2$ is 
constant. Then, equation (\ref{eq:dtilt}) can be reduced to
\begin{eqnarray}
\frac{\partial\vec{l}}{\partial{t}}=\frac{1}{2}\nu_2\frac{\partial^2\vec{l}}{\partial{r}^2}+\vec{T}_{\rm{ex}},
\label{eq:tiltevo}
\end{eqnarray}
where $\vec{l}\cdot\partial\vec{l}/\partial{r}=0$ is used.

%
%
We look for solutions of equation (\ref{eq:tiltevo}) of the form $l_x$, $l_y$$\propto\exp{i}(\omega{t}+kr)$ with $kr\ll1$. Replacing $\partial/\partial{t}$ with $i\omega$, $\partial/\partial{r}$ with $ik$, and $\partial^2/\partial{r^2}$ 
with $-k^2$, we have the following set of linearized equations:
\begin{eqnarray}
\left[
\begin{array}{cccc}
i\omega +\frac{1}{2}\nu_2k^2  && -(\mathcal{A}+ik\mathcal{B}) \\
(\mathcal{A}+ik\mathcal{B})  && i\omega+\frac{1}{2}\nu_2{k^2} \\
\end{array}
\right]
\left(
\begin{array}{cc}
l_x \\
l_y  \\
\end{array}
\right)
=0,
\label{eq:eigen}
\end{eqnarray}
where
\begin{eqnarray}
\mathcal{A}&=&\frac{3}{2}\left(\frac{a}{r}\right)^2
\left[
\frac{1}{2}
\xi_1\xi_2
\Omega
-\frac{\zeta}{\tau_{\rm{rad}}}
\right],
\nonumber \\
\mathcal{B}&=&\Gamma\left[1-\frac{3}{2}\zeta\left(\frac{a}{r}\right)^2
\right].
\nonumber 
\end{eqnarray}
The determinant of {the} coefficient matrix on the left hand side of equation~(\ref{eq:eigen}) 
must vanish because of $\vec{l}\neq0$. The local dispersion relation is then 
obtained as
\begin{eqnarray}
\omega
=
i
\frac{\nu_2k^2}{2}\pm (\mathcal{A}+ik\mathcal{B})
=
i\left\{
\frac{\nu_2k^2}{2}\pm
{k}\Gamma
\left[1-\frac{3}{2}\zeta\left(\frac{a}{r}\right)^2\right]
\right\}
\pm\frac{3}{2}\left(\frac{a}{r}\right)^2\left[
\frac{1}{2}
\xi_1\xi_2
\Omega
-\frac{\zeta}{\tau_{\rm{rad}}}
\right].
\label{eq:circ_omega}
\end{eqnarray}
The imaginary part of $\omega$ corresponds to the excitation or damping of oscillation, 
whereas the real part provides the local precession frequency due to the external torques.

%
%
\begin{figure}[ht!]
\begin{center}
\includegraphics[width=14cm]{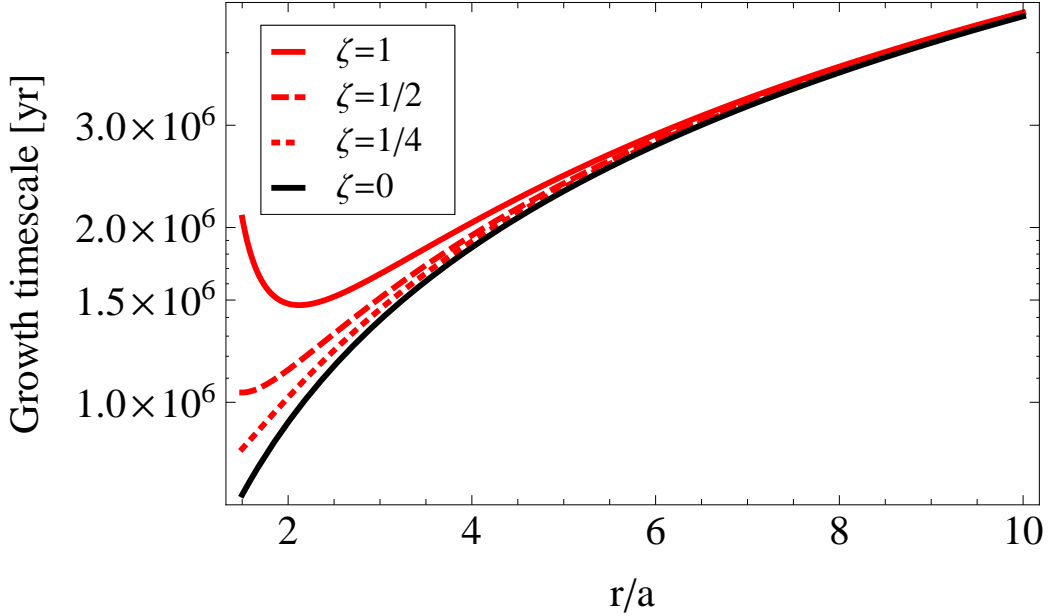}
\end{center}
\caption{Growth timescale of the radiation-driven warping of a circumbinary disk 
with $\alpha=0.1$, $\epsilon=0.1$, $H/r=0.01$, $M=10^7{\rm{M}}_\odot$, and $a=10^4r_{\rm{S}}$. 
The black solid line, red dashed line, and red dotted line show the growth timescales with the binary irradiation parameters $\zeta=0$, $1/4$, $1/2$, and $1$, respectively. The growth timescale with $\zeta=0$ corresponds to {that} of the single black hole case.
}
\label{fig:tgrowth}
\end{figure}

%
%
In order for the perturbation to grow, $\rm{Im}(\omega)$ must be negative. 
The growth condition is given by
\begin{eqnarray}
0<k<\frac{2\Gamma}{\nu_2}\left[1-\frac{3}{2}\zeta\left(\frac{a}{r}\right)^2\right].
\label{eq:waven}
\end{eqnarray}
In terms of $\Gamma_{\rm{bin}}\equiv\Gamma[1-(3/2)\zeta(a/r)^2]$, the growth timescale of the warping mode induced 
by the radiative torques in the binary system is given by
\begin{eqnarray}
\tau_{\rm{rad,bin}}=\frac{r}{\Gamma_{\rm{bin}}}\approx\tau_{\rm{rad}}\left[1+\frac{3}{2}\zeta\left(\frac{a}{r}\right)^2\right].
\label{eq:radt_bin}
\end{eqnarray}
Figure~\ref{fig:tgrowth} shows the dependence of $\tau_{\rm{rad,bin}}$ on binary irradiation parameter 
$\zeta$ and $r/a$ for a model with $\alpha=0.1$, $\epsilon=0.1$, $H/r=0.01$, $M=10^7{\rm{M}}_\odot$, 
and $a=10^4r_{\rm{S}}$. The growth timescale $\tau_{\rm{rad,bin}}$ for $\zeta=0$ or $r/a\ge8$ is reduced 
to the single black hole case.

%
%
We focus our attention on a perturbation with $\lambda\le{r}$, 
where $\lambda=2\pi/k$ is the radial wavelength of the perturbation. 
The condition that the circumbinary disk is unstable to the warping mode 
can be then rewritten as
\begin{eqnarray}
\frac{r}{r_{\rm{S}}}\ge8\pi^2\left(\frac{\eta}{\epsilon}\right)^{2}
\left[1-\frac{3}{2}\zeta\left(\frac{a}{r}\right)^{2}\right]^{-2},
\label{eq:cond}
\end{eqnarray}
where $\eta=\nu_2/\nu_1$ is the ratio of vertical to horizontal viscosities. \cite{og99} derived the relationship between $\eta$ and $\alpha$: $\eta=2(1+7\alpha^2)/(\alpha^2(4+\alpha^2))$ by taking a non-linear effect of the fluid on the warped disk. The value of $\alpha$ consistent with X-ray binary observations is known to be $0.1-0.3$ depending on the state of the accretion disk, although recent magneto-rotational instability simulations provide significantly smaller value of $\alpha$ in a gas-pressure dominated region of the disk (e.g. see \citealt{ob13} and references therein). The range of $\eta$ should therefore be $\eta\gtrsim10$ for $\alpha\lesssim0.3$.
The equality of equation~(\ref{eq:cond}) is approximately held at the marginally stable warping radius:
\begin{eqnarray}
\frac{r_{\rm{warp,bin}}}{r_{\rm{S}}}\approx
\frac{r_{\rm{warp}}}{r_{\rm{S}}}\left[1+3\zeta\left(\frac{a}{r_{\rm{S}}}\right)^{2}\Biggr/\left(\frac{r_{\rm{warp}}}{r_{\rm{S}}}\right)^2\right]
\label{eq:rwb}
\end{eqnarray}
in the case of $r/a>1$ because of $\zeta=\mathcal{O}(0.1)$, where
\begin{eqnarray}
\frac{r_{\rm{warp}}}{r_{\rm{S}}}=8\pi^2\left(\frac{\eta}{\epsilon}\right)^{2},
\label{eq:rws}
\end{eqnarray}
which corresponds to the marginally stable warping radius for a single black hole \citep{pr96}. 
The marginally stable warping radius substantially depends on $\eta$ and the mass-to-energy conversion efficiency $\epsilon$.

\begin{figure}[ht!]
\begin{center}
\resizebox{\hsize}{!}{\includegraphics[]{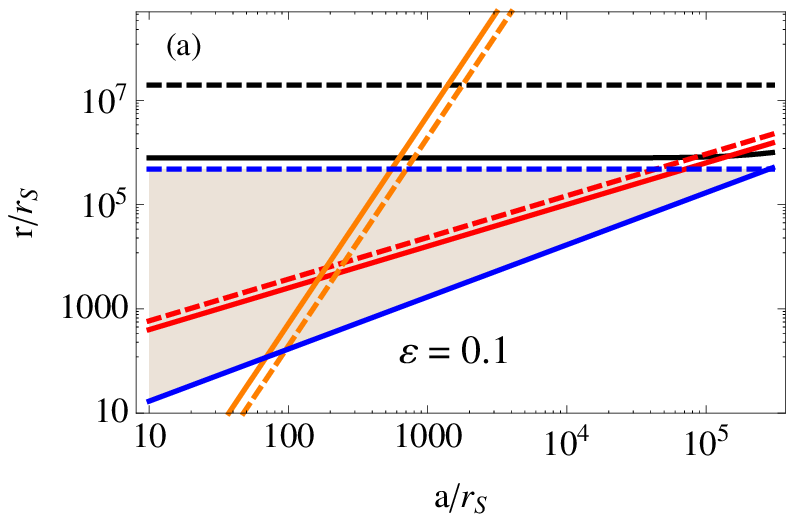}\includegraphics[]{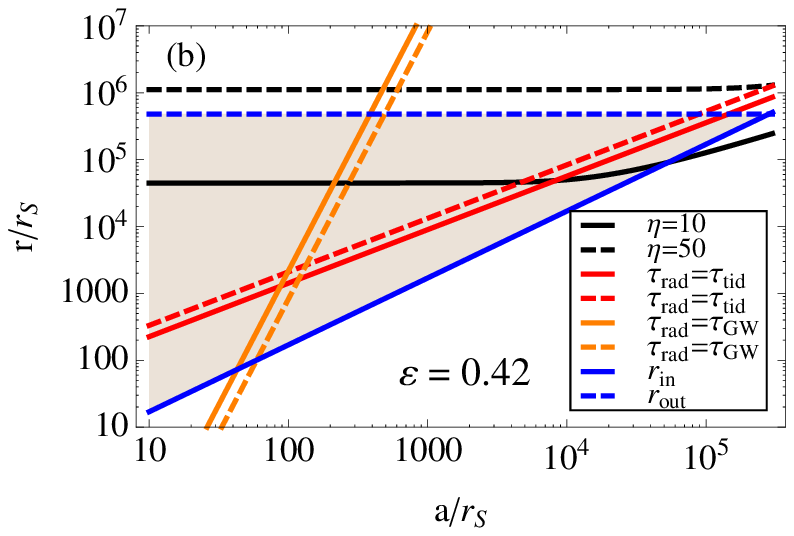}}
\end{center}
\caption{
Characteristic radii of the warped circumbinary disk around binary SMBHs on a circular orbit 
with $\zeta=1$, $q=0.1$, and $M=10^7{\rm{M}}_\odot$. While $\epsilon=0.1$ is adopted in panel~(a), 
$\epsilon=0.42$ is adopted in panel~(b). In both panels, the black solid line and black dashed line 
show the marginally stable warping radii with $\eta=10\,(\alpha=0.10)$ and $\eta=50\,(\alpha=0.27)$, 
respectively. The red lines show the tidal alignment radius where the growth timescale of the radiation-driven 
warping of the circumbinary disk is equal to the timescale during which the circumbinary disk is aligned 
with the orbital plane by the tidal torques. The red solid and dashed lines show the tidal alignment radii 
with $\eta=10$ and $\eta=50$, respectively. The orange lines show the orbital decay radius where the growth timescale of the radiation-driven warping is equal to the orbital decay timescale due to the gravitational 
wave emission. The orange solid and dashed lines show the orbital decay radii with $\eta=10$ and $\eta=50$,
 respectively. While the blue solid line represents the inner radius of the circumbinary disk $r_{\rm{in}}/a\approx1.7$,
  the blue dashed line represents the outer radius of the circumbinary disk $r_{\rm{out}}/r_{\rm{S}}\approx4.8\times10^4\,(M/10^7\,{\rm{M}}_\odot)^{-1/2}$. The shaded area between the blue solid and dashed 
  lines represents the whole region of the circumbinary disk.
}
\label{fig:cradii}
\end{figure}

%
%
Figure~\ref{fig:cradii} shows the dependence of the marginally stable warping radius 
on the semi-major axis. While $\epsilon=0.1$ is adopted in panel~(a), $\epsilon=0.42$ is adopted 
in panel~(b). Panels (a) and (b) thus correspond to the cases of a Schwarzschild black hole and a Kerr black hole with maximum black hole spin parameter, respectively. In both panels, the black solid line and black dashed line show $r_{\rm{warp,bin}}$ normalized by the Schwarzschild radius $r_{\rm{S}}$ for $M=10^7\,{\rm{M}}_\odot$ with $\eta=10\,(\alpha=0.27)$ and $\eta=50\,(\alpha=0.1)$, respectively. The red line shows the radius where the growth timescale of the radiation-driven warping mode, $\tau_{\rm{rad}}$, equals the timescale for the disk to align with the orbital plane by the tidal torque, $\tau_{\rm{tid}}$. This tidal alignment radius is given by
\begin{eqnarray}
\frac{r_{\rm{rad/tid}}}{r_{\rm{S}}}=
\left(\frac{9}{8}\right)^{1/5}
\left(\frac{\xi_1\xi_2}{\epsilon\alpha}\right)^{2/5}
\left(\frac{H}{r}\right)^{-4/5}\left(\frac{a}{r_{\rm{S}}}\right)^{4/5}
\label{eq:rradtid}
\end{eqnarray}
The growth of a finite-amplitude warping mode induced by the radiative torque can be significantly 
suppressed by the tidal torque in the region inside the tidal alignment radius. The red solid 
and dashed lines show the tidal alignment radii with $\eta=10$ and $\eta=50$, respectively. 
The orange line shows the radius where the growth timescale of the radiation-driven warping 
mode equals the timescale in which the binary orbit decays by the gravitational wave emission. 
The orbital decay timescale for a circular binary case is given by \citep{p64}
\begin{eqnarray}
\tau_{\rm{gw}}=\frac{5}{8}\frac{1}{\xi_1\xi_2}\left(\frac{a}{r_{\rm{S}}}\right)^{4}\frac{r_{\rm{S}}}{c}
\label{eq:tgw}
\end{eqnarray}
Equating equation~(\ref{eq:tgw}) with equation~(\ref{eq:trad}), we obtain 
\begin{eqnarray}
\frac{r_{\rm{rad/gw}}}{r_{\rm{S}}}=\frac{5}{32}
\left(\frac{\epsilon\alpha}{\xi_1\xi_2}\right)
\left(\frac{H}{r}\right)^{2}
\left(\frac{a}{r_{\rm{S}}}\right)^{4}.
\label{eq:rradgw}
\end{eqnarray}
Inside this orbital decay radius, the circumbinary disk can be warped 
before two SMBHs coalesce. The orange solid and dashed lines show the orbital decay radii with $\eta=10$ and $\eta=50$, respectively. The blue solid line and blue dashed line show the inner and outer radii of the circumbinary disk, respectively. The inner radius is assumed to be equal to the tidal truncation radius, where the tidal torque is balanced with the viscous torque of the circumbinary disk \citep{al94}. In the case of a circular binary with a small mass ratio, the tidal truncation radius is estimated to be $\sim1.7a$.
 
%
%
A gaseous disk around a SMBH in an AGN is surrounded by a dusty torus. The grains of the dusty torus {are} evaporated above the temperature $1500\,{\rm{K}}$ by the radiation emitted from the central source. The inner radius of the dusty torus should therefore be determined by the dust sublimation radius: $r_{\rm{dust}}=3\,{\rm{pc}}\,(L/10^{46}\,{\rm{erg\,s^{-1}}})^{1/2}(T/1500\,\rm{K})^{-2.8}$, where
$T$ is the dust sublimation temperature \citep{bar87}. Assuming that the AGN luminosity is the Eddington luminosity,
the dust sublimation radius is rewritten as $r_{\rm{dust}}=4.7\times10^{-1}(M/10^7{\rm{M}}_\odot)^{1/2}\,{\rm{pc}}$ with the adoption of $T=1500\,{\rm{K}}$. Since the circumbinary disk should be also inside the dusty torus in our scenario, 
the outer radius of the circumbinary disk is given by
\begin{eqnarray}
\frac{r_{\rm{out}}}{r_{\rm{S}}}\approx4.8\times10^5\left(\frac{M}{10^7{\rm{M}}_\odot}\right)^{-1/2}.
\label{eq:rout}
\end{eqnarray}
The shaded area between the two blue lines shows the whole region of the circumbinary disk.

%
%
%

%
%
It is noted from the figure that the circumbinary disk is not warped by radiation-driven warping
in the cases of $\eta=10$ and $50$ with $\epsilon=0.1$ and $\eta=50$ with $\epsilon=0.42$, 
since the marginally stable warping radii are outside of the circumbinary disk.
On the other hand, the circumbinary disk is warped in the case of $\eta=10$ with $\epsilon=0.42$.
In this case, the marginally stable warping radius corresponds to that of a single black hole at $a/r_{\rm{S}}\lesssim10^4$. If the circumbinary disk around binary SMBHs is warped by radiative torques, the semi-major axis of the binary is predicted to be in a range of $a_{\rm{min}}\lesssim{a}
\lesssim a_{\rm{max}}$, where $a_{\rm{min}}$ is equal to the semi-major axis at the intersection point between $r_{\rm{warp}}$ and $r_{\rm{gw/rad}}$ in panel~(b), which is given as
\begin{eqnarray}
\frac{a_{\rm{min}}}{r_{\rm{S}}}
=
\left[
\frac{32}{5}\left(\frac{\xi_1\xi_2}{\epsilon\alpha}\right)\left(\frac{H}{r}\right)^{-2}\left(\frac{r_{\rm{warp}}}{r_{\rm{S}}}\right)
\right]^{1/4}
\label{eq:amin}
\end{eqnarray}
by equating equation~(\ref{eq:rws}) with equation~(\ref{eq:rradgw}), and $a_{\rm{max}}$ is equal to the semi-major axis at the intersection point between $r_{\rm{rad/tid}}$ and $r_{\rm{out}}$ in panel~(b), which is given as
\begin{eqnarray}
\frac{a_{\rm{max}}}{r_{\rm{S}}}=\left(\frac{8}{9}\right)^{1/4}\left(\frac{\epsilon\alpha}{\xi_1\xi_2}\right)^{1/2}\left(\frac{H}{r}\right)\left(\frac{r_{\rm{out}}}{r_{\rm{S}}}\right)^{5/4}
\label{eq:amax}
\end{eqnarray}
by equating equation (\ref{eq:rradtid}) with equation~(\ref{eq:rout}).

%
%
\begin{figure}[ht!]
\begin{center}
\includegraphics[width=14cm]{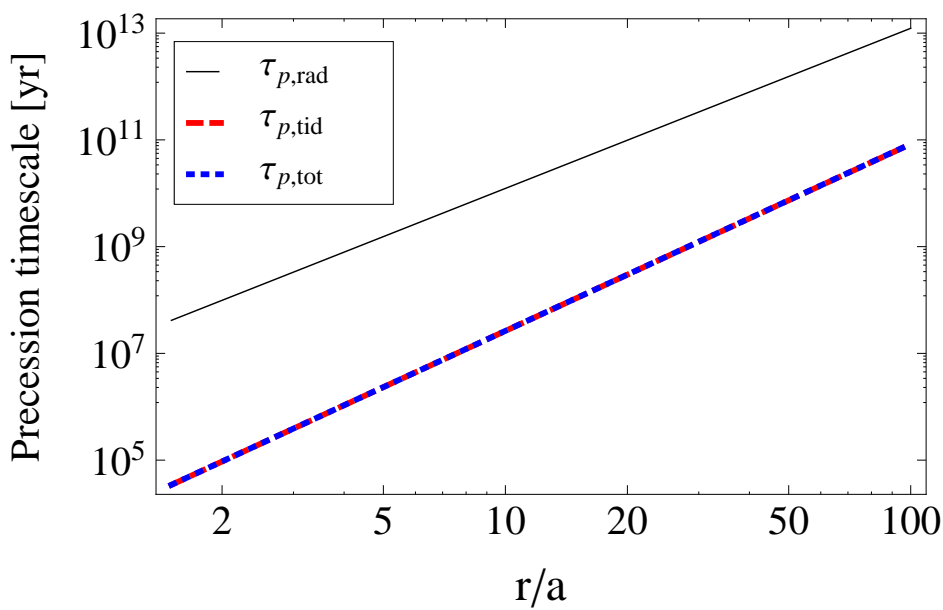}
\end{center}
\caption{Precession timescale of the warped circumbinary disk around binary SMBHs 
on a circular orbit with $\zeta=1$, $q=0.1$, $\epsilon=0.1$, $\alpha=0.1$, $H/r=0.01$, and $M=10^7{\rm{M}}_\odot$. 
The black solid line, blue dotted line, and red dashed line show the precession timescales {from} the 
radiative torque, {from} the tidal torque, and {from} the sum of those two torques, respectively.
}
\label{fig:precess}
\end{figure}

%
%
The local precession frequency {$\Omega_{\rm{p,tot}}$} of the linear warping mode 
is obtained from equation~(\ref{eq:circ_omega}) by
\begin{equation}
\Omega_{\rm{p,tot}}={\rm Re}(\omega)=-\frac{3}{2}\left(\frac{a}{r}\right)^2\left[
\frac{1}{2}
\xi_1\xi_2
\Omega
-\frac{\zeta}{\tau_{\rm{rad}}}
\right]=-(\Omega_{\rm{p},\rm{tid}}-\Omega_{\rm{p,rad}}),
\label{eq:pfreq}
\end{equation}
where 
\begin{eqnarray}
\Omega_{\rm{p},\rm{tid}}&=&\frac{3}{4}\left(\frac{a}{r}\right)^2
\xi_1\xi_2
\Omega,
\nonumber \\
\Omega_{\rm{p,rad}}&=&\frac{3}{2}\left(\frac{a}{r}\right)^2\frac{\zeta}{\tau_{\rm{rad}}}.
\nonumber
\end{eqnarray}
The radius where $\Omega_{\rm{p},\rm{tid}}$ is balanced 
with $\Omega_{\rm{p},\rm{rad}}$ is given by
\begin{eqnarray}
\frac{r_{\Omega_{\rm{rad/tid}}}}{r_{\rm{S}}}
\sim1.3\times10^{11}
\left(
\frac{\xi_1\xi_2}{1/4}
\right)^{2}
\left(\frac{0.1}{\epsilon}\right)^2
\left(\frac{0.1}{\alpha}\right)^2\left(\frac{1}{\zeta}\right)^2\left(\frac{H/r}{0.01}\right)^4.
\end{eqnarray}
Since $r_{\Omega_{\rm{rad/tid}}}\gg{r}_{\rm{out}}$, the tidal precession frequency is {higher} than the radiative precession frequency. Thus, the circumbinary disk slowly precesses in the retrograde direction.

%
%
Figure~\ref{fig:precess} shows the dependence of the 
precession timescales on the circumbinary disk radius normalized by the semi-major axis. 
The black solid line, the red dashed line, and the blue dotted line show the precession timescale for the radiative torques  $\tau_{\rm{p,rad}}=1/\Omega_{\rm{p,rad}}$, tidal torques $\tau_{\rm{p,tid}}=1/\Omega_{\rm{p,tid}}$, and total torque $\tau_{\rm{p,tot}}=1/\Omega_{\rm{p,tot}}$, respectively. The precession timescale is much longer than the orbital period. 

%
\section{Application to observed maser disks in AGNs}
\label{sec:4}
%

%
%
In this section, we discuss the application of our proposed model 
to a warped maser disk system. There {is} observational {evidence} 
for disk warping in the maser disks at the center of NGC 4258 \citep{he05}, 
Circinus \citep{green03}, NGC\,2273, UGC\,3789, NGC\,6264, and NGC\,6323 \citep{kuo11}.
We assume that these maser disks start to be warped at the innermost maser spot radii, 
which we call the observed warping radii, in the following discussion.

%
%
From equation~(\ref{eq:rws}), 
the marginally stable warping radius for an extreme Kerr black hole with $\eta=10$ is estimated to be
\begin{eqnarray}
\frac{r_{\rm{warp}}}{r_{\rm{S}}}\sim4.4\times10^{4}\left(\frac{\eta}{10}\right)^2
\left(\frac{0.42}{\epsilon}\right)^2.
\label{eq:rwarp}
\end{eqnarray}
Equation (\ref{eq:rwarp}) is a good approximation to the marginally stable warping radius {for} a binary 
SMBH as long as $a/r_{\rm{S}}\lesssim10^4$, as seen in the solid black line at panel~(b) of 
Figure~\ref{fig:cradii}. In order for the maser disks to be warped, the marginally stable warping radius 
must be less than not only the outer radius of the circumbinary disk but also the observed warping radius. 
Otherwise, radiation-driven warping is unlikely {as} a mechanism to explain the warped structure of these maser disks. We adopt this condition in order to examine whether our model is appropriate for the observed warped maser disks.

%
%
Table~\ref{tbl:t2} summarizes the results of {applying} our model 
to observed warped maser disks. 
The first and second columns denote the name and 
observed black hole mass of each target system, respectively. 
The third and fourth columns represent 
the observed warping radius, $r_{\rm{warp}}^{\rm{obs}}$ and 
outer radius of the circumbinary disk, respectively. 
The outer radius is obtained by equation~(\ref{eq:rout}).
The fifth to sixth columns denotes the 
inferred semi-major axis for each target system, if the observed warped maser disk is a circumbinary disk 
around binary SMBHs and the observed warping radius is larger than the marginally stable warping radius 
given by equation~(\ref{eq:rwarp}). Since their observed warping radii also are smaller than the outer radius 
of the circumbinary disk, they intersect with two lines of $r_{\rm{rad/tid}}$ and $r_{\rm{rad/gw}}$, respectively.
It is clear from panel~(b) of Figure~\ref{fig:cradii} that the semi-major axis at the intersection point 
between $r_{\rm{warp}}^{\rm{obs}}$ and $r_{\rm{rad/tid}}$
provides the maximum value of the inferred semi-major axis, 
whereas the semi-major axis at the intersection point between $r_{\rm{warp}}^{\rm{obs}}$ and $r_{\rm{rad/gw}}$ 
gives the minimum value of the inferred semi-major axis.
Each semi-major axis is then obtained by equating each observed warping radius with
 equations~(\ref{eq:rradtid}) and (\ref{eq:rradgw}) as
\begin{eqnarray}
\frac{a_{\rm{min}}^{\rm{obs}}}{r_{\rm{S}}}
&=&
\left[
\frac{32}{5}\left(\frac{\xi_1\xi_2}{\epsilon\alpha}\right)\left(\frac{H}{r}\right)^{-2}\left(\frac{r_{\rm{warp}}^{\rm{obs}}}{r_{\rm{S}}}\right)
\right]^{1/4},
\label{eq:semin}
\\
\frac{a_{\rm{max}}^{\rm{obs}}}{r_{\rm{S}}}
&=&\left(\frac{8}{9}\right)^{1/4}\left(\frac{\epsilon\alpha}{\xi_1\xi_2}\right)^{1/2}\left(\frac{H}{r}\right)\left(\frac{r_{\rm{warp}}^{\rm{obs}}}{r_{\rm{S}}}\right)^{5/4},
\label{eq:semimax}
\end{eqnarray}
where we adopt that $H/r=0.01$ and $q=0.1\,(\xi_1\xi_2=10/121)$. The corresponding orbital periods are shown 
in the seventh and eighth columns, respectively. 

%
%
From Table~\ref{tbl:t2}, {only the Circinus meets the condition that the observed warping radius is larger than the marginally stable warping radius for $\eta=40$ and $\epsilon=0.42$, {while being} smaller than the outer radius of the circumbinary disk. On the other hand, all systems, except for NGC\,2273, satisfy the same condition but for $\eta=10$ and $\epsilon=0.42$.}
The radiation-driven warping can thus be a promising mechanism for explaining the warped structure of the observed maser disks in these systems. There is also a possibility that the central massive objects are binary SMBHs with the semi-major axis on several tens of milliparsec to sub-milliparsec scales. 
However, it is difficult to distinguish, solely by the current analysis, whether the central object is a single SMBH or binary SMBHs. To do so, independent theoretical and observational approaches are needed.

%
%
\begin{landscape}
 \begin{table*}[!h]
  \caption{
Application to observed warped maser disks. The first column denotes the name of each target system. The second and third columns show the black hole mass and innermost-maser-spot radius of each target system, respectively (see \cite{green03,he05,kuo11,kho13}). The fourth column represents the outer radius of the circumbinary disk, which is given by equation~(\ref{eq:rout}). The fifth and sixth columns denote the semi-major axes estimated by 
equations~(\ref{eq:semin}) and (\ref{eq:semimax}), respectively. The seventh and final columns indicate the corresponding orbital periods. {Note that the marginally stable warping radii are estimated by equation~(\ref{eq:rwarp}) to be $4.4\times10^4\,r_{\rm{S}}$ for $(\eta,\epsilon)=(10,0.42)$ and $7.2\times10^5\,r_{\rm{S}}$ for $(\eta,\epsilon)=(40,0.42)$, respectively.}
  }
   \begin{center}
     \begin{tabular}{cccccccc}
       \hline \hline
         Target system & $M\,[{\rm{M}}_\odot]$ 
         & $r_{\rm{warp}}^{\rm{obs}}\,[r_{\rm{S}}]$
         & $r_{\rm{out}}\,[r_{\rm{S}}]$ 
         & $a_{\rm{min}}^{\rm{obs}}\,[r_{\rm{S}}]$ 
         & $a_{\rm{max}}^{\rm{obs}}\,[r_{\rm{S}}]$ 
          & $P_{\rm{min}}^{\rm{obs}}\,[{\rm{yr}}]$ 
         & $P_{\rm{max}}^{\rm{obs}}\,[{\rm{yr}}]$ \\
       \hline \hline
       NGC\,4258 & $3.78\times10^7$ & $4.70\times10^{4}$ & $2.48\times10^5$ & $216$ & $7.92\times10^3$ & $3.32\times10^{-1}$ & $73.9$\\
       Circinus      & $1.14\times10^6$ & $1.01\times10^{6}$  & $1.43\times10^6$ & $464$ & $3.66\times10^5$ & $3.16\times10^{-2}$ & $699$\\
       NGC\,2273 & $7.50\times10^6$ & $3.90\times10^{4}$  & $5.56\times10^5$ & $-$ & $-$ & $-$ & $-$\\
       UGC\,3789 & $1.04\times10^7$ & $8.44\times10^{4}$  & $4.72\times10^5$ & $250$ & $1.65\times10^4$ & $1.14\times10^{-1}$ & $60.9$\\
       NGC\,6264 & $2.91\times10^7$ & $8.62\times10^{4}$  & $2.82\times10^5$ & $251$ & $1.69\times10^4$ & $3.20\times10^{-1}$ & $177$\\
       NGC\,6323 & $9.40\times10^6$ & $1.45\times10^{5}$  & $4.97\times10^5$ & $286$ & $3.22\times10^4$ & $1.26\times10^{-1}$ & $151$\\
       \hline
     \end{tabular}
   \label{tbl:t2}
   \end{center}
 \end{table*}
 \end{landscape}

%
\section{Summary and Discussion}
\label{sec:5}
%
%
We have investigated {the} instability of a warping mode in a geometrically thin, non-self-gravitating 
circumbinary disk induced by radiative torques originated from two accretion disks around interior black holes. Here, {the two accretion disks} are regarded as point {irradiation sources} for simplicity. We have derived the condition where the circumbinary disk is unstable to the warping mode induced by the radiative torques and the timescales of precession caused by both tidal and radiative torques for a small tilt angle {($\beta\ll1$)}. Our main conclusions other than {this} instability condition are summarized as follows:
\begin{enumerate}
\renewcommand{\theenumi}{\arabic{enumi}}
\item For $r/a\gtrsim{8}$, the growth timescale of the warping mode in the binary SMBH case 
is reduced to that of the {single} SMBH case.
\item The marginally stable warping radius substantially depends on both the ratio of the vertical to horizontal shear viscosities $\eta$ and the mass-to-energy conversion efficiency $\epsilon$. The marginally stable warping radius 
in the binary SMBH case is reduced to that of the single SMBH case for $r\gg{a}$.
\item For a small tilt angle ($\beta\ll1$), the tidal torques due to the binary potential give no contribution to the growth of warping modes on the circumbinary disk.
\item There is a clear difference in the warping radius between the single SMBH case and the binary SMBHs case. Since the tidal torques work on the circumbinary disk so as to align the circumbinary disk plane with the binary orbital plane, they can suppress finite-amplitude warping modes induced by the radiative torques. The circumbinary disk, therefore, starts to be warped at the tidal alignment radius where the growth timescale of the radiation-driven warping of the circumbinary disk is equal to the timescale for which the disk is aligned with the orbital plane by the tidal torques, if the tidal alignment radius is larger than the marginally stable warping radius. In contrast, the accretion disk around a single SMBH starts to be warped at the marginally stable warping radius.
\item If the circumbinary disk is warped by radiative torques due to radiation emitted from two accretion disks
around the black holes, the binary SMBHs with masses of  $10^{7}{\rm{M}}_\odot$ are likely to have a binary 
separation on $10^{-2}\,\rm{pc}$ to $10^{-4}\,\rm{pc}$ scales.
\item The circumbinary disk can precess due to both tidal torques and radiative torques. 
While the radiative torques tend to precess the circumbinary disk in the prograde direction, 
the tidal torques tend to precess it in the retrograde direction. Since the former precession 
frequency is much lower than the latter precession frequency, the circumbinary disk slowly 
precesses in the retrograde direction. The precession timescale is much longer than the orbital 
period. Therefore, it is unlikely {that} the periodic light variation due to the 
warped precession {could be detected.}
\end{enumerate}

%
%
In this paper, we have studied warping of circumbinary disks where disk self-gravity is negligible.
A few warped maser disks are, however, thought to be massive to be comparable to the black hole 
mass (e.g., \citealt{wy12}). {The self-gravitating force in such a massive disk} makes the velocity profile 
{deviate significantly} from the Keplerian one. In addition, the dominant origin of both the horizontal and 
vertical shear viscosities, on which the condition of the radiation-driven warping is sensitive, is the 
self-gravitating instability of the disk. However, little is known about how the self-gravitating force 
affects disk warping in a geometrically thin, self-gravitating circumbinary disk consistent with the maser 
disk observations. Further observational and theoretical studies are necessary.

%
%
We have assumed that the binary is on a circular orbit. There are, however, 
theoretical indication that the orbital eccentricity increases by the interaction between binary SMBHs and their circumbinary disks \citep{an05,kh09}. 
In the ideally efficient binary-disk interaction case, the orbital eccentricity is driven up to $\sim0.6$. 
This is because the binary orbital angular momentum is mainly transferred to the circumbinary disk 
when the binary is at the apastron. The saturation value of the orbital eccentricity is estimated by 
equating the angular frequency at the inner radius of the circumbinary disk with the binary orbital 
frequency at the apastron \citep{hui10,rdscc11}. In addition, more enhanced periodic light variations 
are expected in eccentric binary SMBHs by interaction with the circumbinary disk \citep{hms07,hmh08} 
than in the circular binary case \citep{mm08,dza13}. Such periodic light curves provide {an independent tool} to evaluate whether the central object of the warped maser disk is binary SMBHs or a single SMBH. We will examine the effect of the orbital eccentricity on the radiation-driven warping of the circumbinary disk in a subsequent paper.

%
%
{
For simplicity, we have also assumed that the circumbinary disk is initially aligned with 
the binary orbital plane ($\beta\ll1$), as in most of the previous studies. However, the 
angular momentum vector of the circumbinary disk does not always coincide with that 
of the binary orbital angular momentum, because the orientation of the circumbinary disk 
is primarily due to the angular momentum distribution of the gas supplied to the central 
region of AGNs. Therefore, the orientation of the circumbinary disk plane can be taken 
arbitrarily with respect to the binary orbital plane. In such a misaligned system with a 
significant tilt angle, the inner part of the circumbinary disk tends to align with the binary 
orbital plane by the tidal interaction between the binary and the circumbinary disk, whereas 
the outer part tends to retain the original state by the shear viscosity in the vertical direction. 
As a result, the circumbinary disk should be warped without the effect of radiation driven 
warping instability. It is important to examine how the radiation driven warping instability 
works in the misaligned systems under the tidal potential, but it is difficult to find the analytic 
solutions because of the complicated dependence of  the tidal and radiative torques on the 
tilt angle and azimuth of tilt. We will numerically study this problem in the future.
}

%
%
There is a cavity between the circumbinary disk and the binary (see Figure~\ref{fig:schmatic}), 
which is elongated even in a circular binary case because of the binary-disk interaction (e.g., \citealt{mm08}). 
The inner radius of the circumbinary disk, i.e., the outer radius of the cavity, is equal to the tidal truncation 
radius, where the tidal torque is balanced with the viscous torque of the circumbinary disk, and is typically 
$\sim 2a$. Since the marginally stable warping radius is substantially larger than the inner radius of the 
circumbinary disk, the shape of the cavity gives little influence on the warping condition.

%
%
Probing gravitational waves (GWs) from individual binary SMBHs with masses $\gtrsim10^7{\rm{M}}_\odot$ 
with Pulser Timing Arrays (PTAs) \citep{lb01,svv09} also gives a powerful tool to determine if the central 
object surrounded by the warped maser disk is binary SMBHs or a single SMBH. 
For a typical PTA error box ($\approx40\,\rm{deg}^2$) in the sky, the number of interloping AGNs are of the order of $10^2$ for more than $10^8\,{\rm M}_\odot$ black holes if the redshift range is between $0$ and $0.8$ (see Figure 1 of \citealt{tk12} in detail). Assuming that the central objects at the center of observed warped maser disks are binary SMBHs on several tens of milliparsec scale, the characteristic amplitudes of GWs emitted from those systems are estimated to be $\lesssim10^{-17}$ for inspiral GWs and $\lesssim10^{-16}$ for memory GWs associated with the final mergers \citep{seto09}. Since they are three to four orders of magnitude less than the current PTA sensitivity of $\gtrsim10^{-13}$, it is unlikely for GWs to be detected from 
the currently identified warped maser disk systems. If the total mass of binary SMBHs is more massive, however, 
the characteristic amplitudes of the GW signals could be large enough to be detected with future planned 
PTAs such as the Square Kilometer Array with $\gtrsim10^{-16}$ sensitivity. It will therefore be desired to 
identify warped maser disks around the central massive objects with masses $\gtrsim10^8{\rm{M}}_\odot$ 
in nearby AGNs.

%
%
We have also discussed the application of the warped circumbinary disk model to the observed warped maser 
disks in Table~\ref{tbl:t2}. In the case of the marginally stable warping radius with $\eta=40\,(\alpha\approx0.1)$ 
and $\epsilon=0.42$, only the Circinus meet{s} the condition that the marginally stable warping radius {is} 
less than both the observed warping radius and the dust sublimation radius of AGN which is assumed to be equal to 
the outer radius of the circumbinary disk. In this case, the resultant inferred semi-major axis is between 
$6.3\times10^{-5}\,\rm{pc}$ and $2.6\times10^{-2}\,\rm{pc}$. On the other hand, it is unlikely that the warped 
structure of the maser disks at the center of other five systems {originates from radiative torque}, even 
{if} their central objects are a single SMBH. The condition in question substantially depends on the observed 
warping radius and values of $\eta$ and $\epsilon$ in the marginally stable warping radius. Further theoretical 
arguments about an appropriate treatment of $\epsilon$ and $\eta$, and observations to measure the warping radii 
more precisely in the existing maser disks, are desirable.

%
\section*{Acknowledgments}
%
{
The authors thank the anonymous referee for fruitful comments and suggestions. 
The authors also thank Nicholas Stone for his carefully reading the manuscript 
and helpful comments.} KH is grateful to Jongsoo~Kim for helpful discussions 
and his continuous encouragement. BWS and THJ are grateful for support from 
KASI-Yonsei DRC program of Korea Research Council of Fundamental Science 
and Technology (DRC-12-2-KASI). This work was also supported in part by the 
Grants-in-Aid for Scientific Research (C) of Japan Society for the Promotion of 
Science [23540271 TN and KH, 24540235 ATO and KH].

%
%
%

\end{document}